\documentclass[pdftex,aps,prl,10pt,twocolumn,nofootinbib,showpacs,showkeys,notitlepage]{revtex4-1}
\usepackage{amsmath,graphicx,bm}
\pdfoutput=1

\renewcommand{\d}{\partial}

\newcommand{\nn}{\nonumber\\}

\newcommand{\ph}{\varphi}
\newcommand{\rh}{\varrho}
\newcommand{\exv}[1]{\left\langle{#1}\right\rangle}

\newcommand{\ep}{\varepsilon}

\renewcommand{\k}{{\bf k}}
\newcommand{\p}{{\bf p}}

\newcommand{\sgn}{\mathop{\textrm{sgn}}}

\renewcommand{\Re}{\,\textrm{Re}\,}
\newcommand{\pint}[2]{{\int\!\frac{d^{#1}#2}{(2\pi)^#1}\,}}
\newcommand{\pintz}[1]{{\int\!\frac{d #1}{2\pi}\,}}

\newcommand\lsim{\mathrel{\rlap{\lower4pt\hbox{\hskip1pt$\sim$}} \raise1pt\hbox{$<$}}}                
\newcommand\gsim{\mathrel{\rlap{\lower4pt\hbox{\hskip1pt$\sim$}} \raise1pt\hbox{$>$}}}                

\newcommand{\K}{{\cal K}}

\begin{document}

\title{Non-particle statistical physics}

\author{A. Jakov\'ac}
\email{jakovac@phy.bme.hu}
\affiliation{Dept. of Theoretical Physics, BME Technical University, 
H-1111 Budapest, Hungary}

\date{\today}

\begin{abstract}
  Consistent statistical physical description is given for systems
  where the elementary excitations are composite objects. Explicit
  calculational scheme is constructed for the energy density and the
  total number of thermodynamical degrees of freedom, based on the
  spectral function of the system. One demonstrates through
  characteristic examples that single quasiparticle contributions
  combine non-linearly in these quantities. Relation to the Gibbs
  paradox is also discussed.
\end{abstract}

\maketitle

QCD thermodynamics shows paradoxical faces in the crossover region
connecting the hadronic and the quark phases. Variants of the
perturbative approach, which provide a reasonable interpretation of
the numerical simulations in restricted temperature ranges, work with
perturbed ideal gas systems where the mean free path ($\xi$) is very
large. On the other hand, as for example the interpretation of recent
RHIC measurements suggest \cite{Shuryak:2003xe,Teaney:2003kp},
hadronic matter near $T_c$ behaves as an almost ideal fluid where
$\xi\to0$. Enforcing a free gas description to an ideal fluid,
however, is possible only with strong interactions.

But what seems to be strongly interacting in one way, can be weakly
interacting in terms of the adequate degrees of freedom. In PT one
uses resummation to find them; and indeed, perturbative series are
improved considerably eg. for entropy with HTL resummation
\cite{Blaizot:1999ip}, thermodynamics with screened PT
\cite{Karsch:1997gj} or with 2PI resummation \cite{Berges:2004hn}. In
effective approaches like the hadron resonance gas model or massive gluon
picture, also \emph{noninteracting} particles can give an account for
complicated behavior in QCD like charm hadronization
\cite{Andronic:2003}, equation of state
\cite{Karschetal,Huovinen:2009yb, Borsanyi:2010cj}, or interaction
measure \cite{Castorina:2011ja}. The success of these approaches
suggests that probably most part of the strong interactions between
free particles is incorporated in the \emph{spectrum}, and the
residual interactions of the so-dressed excitations are small.

We therefore may assume that the adequate description is based on a
model which contains weakly interacting excitations with nontrivial
spectrum (not necessarily quasiparticles). In this paper we consider
the most simple effective theory described by a single real bosonic
excitation, and omit all interactions. We primarily interested in the
questions of how finite lifetime or multiparticle (threshold) effects
can modify the thermodynamics, and how can different degrees of
freedom show up in or vanish from the statistical ensemble
dynamically.

The treatment of an effective field theory with a quite general
spectrum, however, requires to rethink some relevant questions
concerning the foundations of statistical physics. The problem is the
following: the partition function
\begin{equation}
  \label{Z}
  Z = \sum_n g_n e^{-\beta E_n}
\end{equation}
requires the knowledge of the energy levels $E_n$ as well as the
multiplicities $g_n$ (for simplicity we assume no conserved charge,
ie. no chemical potential here). In a fundamental theory the basic
excitations are elementary, therefore their multiplicities are fixed
permanently, while in an effective theory the thermodynamic degrees of
freedom might change dynamically. To have a cleaner characterization
of the multiplicities, and also to approach the question of
appearance/disappearance of modes in the ensemble, let us define the
total number of (internal) degrees of freedom as the sum of
degeneracies in a 1-particle case at zero spatial momentum
\begin{equation}
  N_{dof} = \sum_n g_n \biggr|_{1-particle, \k=0}.
\end{equation}
The value of $N_{dof}$ is always an integer in the fundamental theory
and it appears directly in physical quantities: for example the
pressure of bosonic gases at high temperature is $P=N_{dof} P_{SB}$,
where $P_{SB}$ is the Stefan-Boltzmann (SB) limit $P_{SB}=\pi T^4/90$.

\begin{figure}[htbp]
  \centering
  \hbox{\hspace*{-2em}\includegraphics[height=3.8cm]{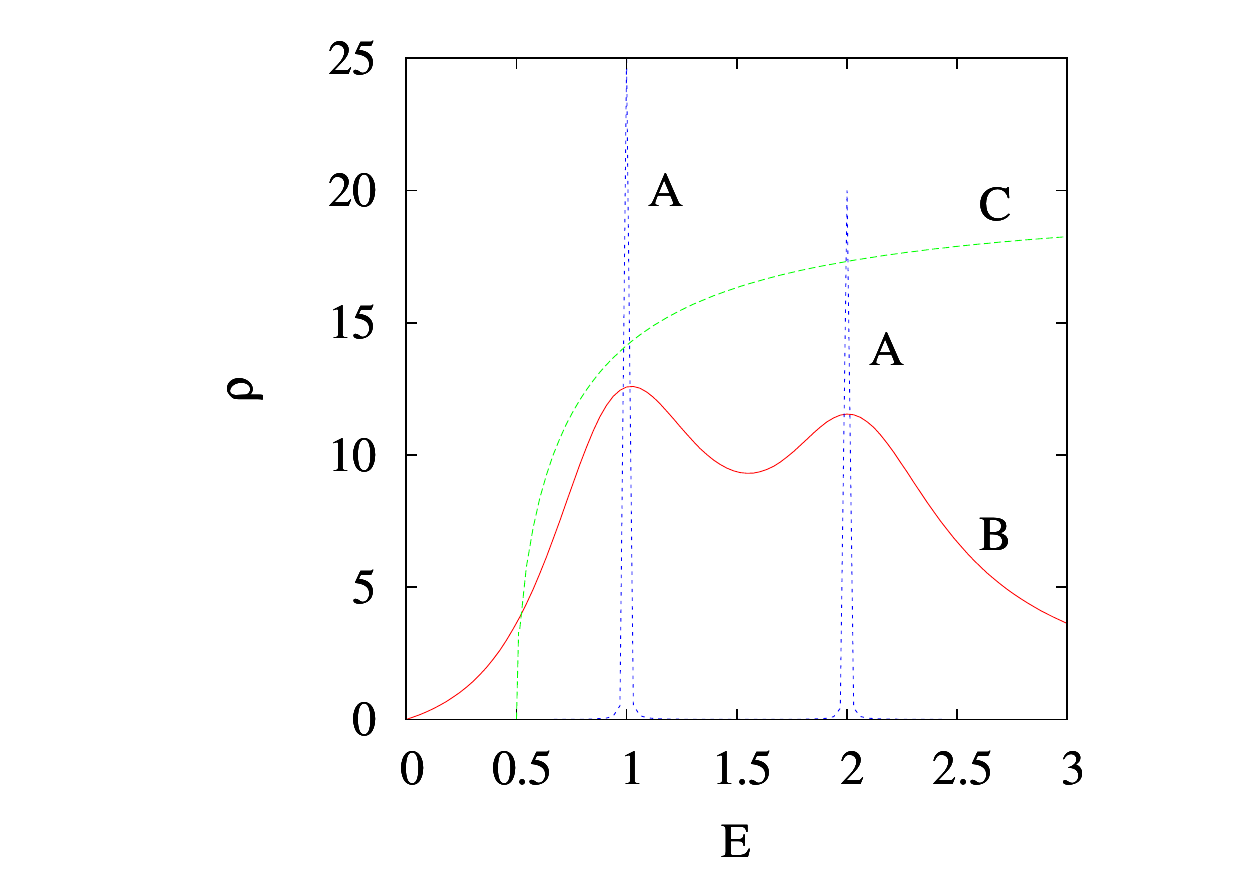}
    \hspace*{-4em}\includegraphics[height=3.8cm]{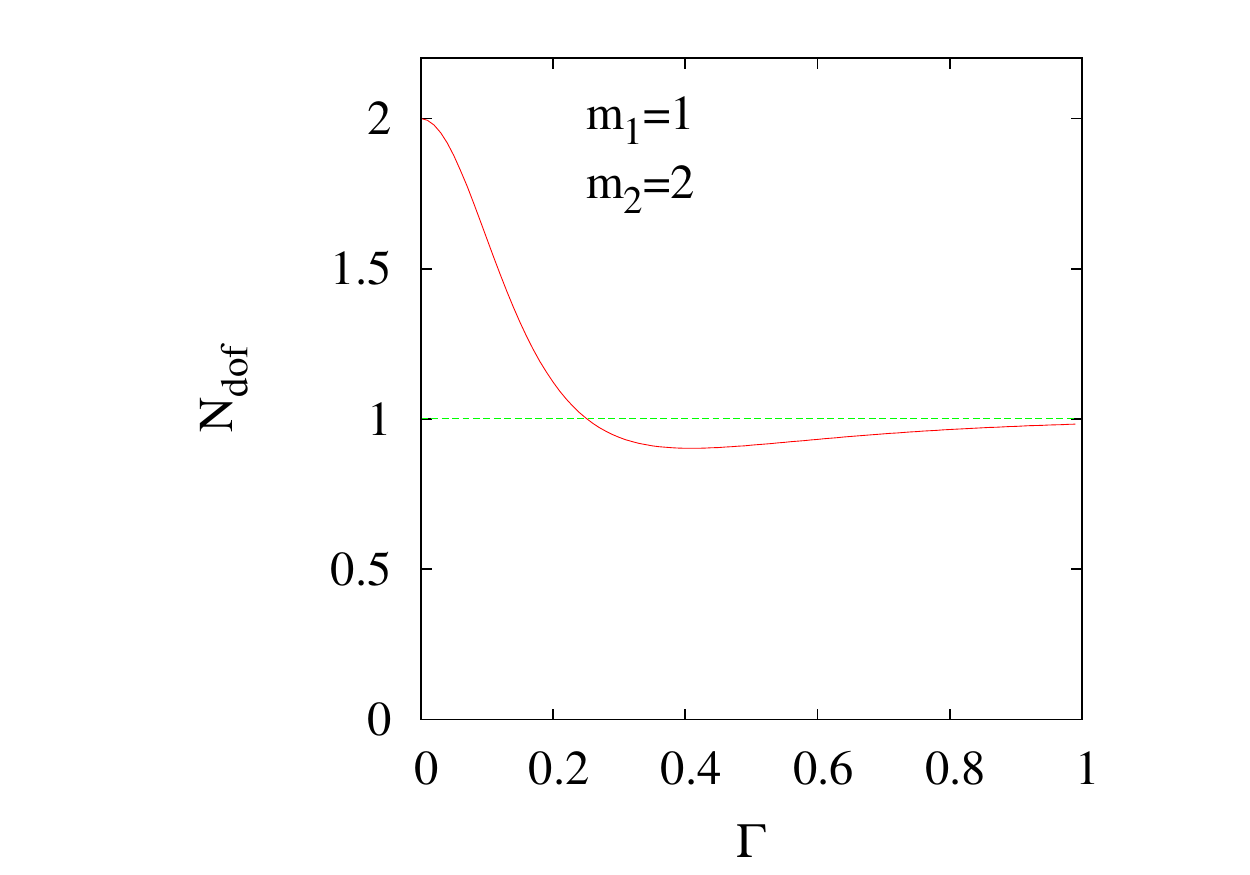}}
  \hbox{\hspace*{7em}a.)\hspace*{12em}b.)}
  \caption{a.) spectral function for: A.) well separated peaks, B.)
    merging peaks and C.) threshold. Figure b.) shows the behavior of
    the degrees of freedom from \eqref{np} as we go from case A to
    case B (and beyond).}
  \label{fig:twostates}
\end{figure}

In an effective theory, however, elementary excitations are in fact
multiparticle states from the point of view of the underlying
fundamental theory. Fig.~\ref{fig:twostates}/a shows three
characteristic cases how the multiparticle nature manifests itself in
the spectral function $\rh$.
In case A the two well-separated peaks each represent independent
quasiparticles: they can be treated as a fundamental particles, and so
$N_{dof}=2$ here. In case B we should avoid the double-counting of the
states in the overlap region, so we expect $N_{dof}<2$. If the widths
$\Gamma\to\infty$, finally we get one peak and we expect
$N_{dof}\to1$. The behavior of $N_{dof}$ as a function of the widths
should be something like Fig.~\ref{fig:twostates}/b -- which results
actually from an explicit evaluation (cf. below \eqref{np}). Case
\emph{C} is even more difficult, since there is no quasiparticle-like
excitation at all; but we have to determine also its contribution to
the thermodynamics.

The problem of determining the correct degeneracies is in close
connection with the old problem of indistinguishability of particles
in gases and the Gibbs paradox. Consider two energy levels with
separation $\Delta E$, then $N_{dof}=2$ for $\Delta E>0$, even for
$\Delta E\to0$, but $N_{dof}=1$ if $\Delta E=0$ (since then we have
only one level). Therefore $N_{dof}$, and so the SB limit of the
energy density is a non-analytic function of $\Delta E$. However,
according to the argumentation of the previous paragraph, for real
gases with finite (eg. thermal) width, for $\Delta E<\Gamma$ the
number of degrees of freedom $N_{dof}$ goes smoothly to one. A similar
conclusion was drawn in \cite{GibbsQM}. By a better description of the
degeneracies in effective theories we may hope to have a deeper
understanding of the Gibbs paradox in general.

The key observation for the correct description is that statistical
physics is determined by the dynamics, so we shall start with the
Lagrangian of the effective theory. In case of a quadratic model the
spectral function fixes completely the dynamics, and so the
thermodynamics and also $N_{dof}$. Below the spectral function will be
the input of the analysis and the consequence of a few physically
appealing non-trivial choices will be explored.

The most general quadratic, spacetime translation invariant action
reads
\begin{equation}
  \label{eq:action}
  S[\ph] = \frac12 \int d^4xd^4y\,\ph(x) K(x-y)\,\ph(y).
\end{equation}
The kernel $K$ is to be constructed to represent the given spectral
function. If it is continuous, $K$ should be time-nonlocal. This
immediately poses the question of causality; but causality in this
system simply means that the spectral function is zero for spacelike
separation. This requirement is granted by construction, since the
spectral function serves as input here.

We should note that because of time-nonlocality some standard
techniques will not work. First of all we do not have canonical
formalism (or at least we would need infinitely many fields). Also
there is no direct connection between the imaginary time formalism
and the thermodynamics: indeed, the standard derivation is based on
the fact that the Hamiltonian density depends on the canonical
momentum as $\Pi^2/2$ \cite{LeBellac}. In order to define
thermodynamics we have to define the energy through the Noether current
belonging to the time translation invariance, and then we should
compute its expectation value at finite temperatures. We note that the
formal correspondence between the time translation $e^{-iHt}$ and
statistical operator $e^{-\beta H}$ is still valid, and so KMS
condition remains true: this will help us to take expectation
values. Once we know the temperature dependent energy $E(T)$, we
can define the complete thermodynamics from that.

To connect the kernel to the spectral function we compute the analytic
continuation of the retarded/advanced propagator from the spectral
function by the Kramers-Kronig relation
\begin{equation}
  G_R(p) = \pintz\omega \frac{\rh(\omega,\p)}{p_0-\omega+i\epsilon}.
\end{equation}
The kernel in Fourier space (denoted by $\K$), can be obtained from
the relation: $G_R(p) = \K^{-1}(p_0+i\epsilon,\p)$. Using the
hermiticity of the action \eqref{eq:action}, there remains
\begin{equation}
  \K(p) = \Re G^{-1}_R(p).
\end{equation}
If the spectral function is relativistic invariant, then the same
holds for $G_R$ and for the kernel, too. From now on we assume
relativistic invariance (although for the applicability of the
approach it is not necessary).

The next step is the construction of the energy-momentum tensor: it is
the conserved current belonging to the spacetime translation
symmetry. Using standard techniques \cite{PeskinSchroeder} we obtain
\begin{equation}
  \hat T_{\mu\nu}(x) = \frac12 \ph(x)\,D_{\mu\nu}\K(i\d)\,
  \ph(x),
\end{equation}
where
\begin{equation}
  D_{\mu\nu}\K(p) = p_\mu\frac{\d\K}{\d p^\nu}\, - g_{\mu\nu}\K =  2
  p_\mu p_\nu \frac{d\K}{d p^2}\, - g_{\mu\nu}\K.
\end{equation}
The latter, relativistic form is explicitly symmetric.

Expectation value of this form can be taken using the KMS condition
$iG^{12}(p)= n(p_0)\rh(p)$, where $n$ is the Bose-Einstein
distribution. We find
\begin{equation}
  \exv{T_{\mu\nu}} = \int\limits_p^+ D_{\mu\nu}\K(p)\,
  \left(\frac12+n(p_0)\right)\rh(p),
\end{equation}
where $\int_p^+ = \int_0^\infty \frac{dp_0}{2\pi}\pint3\p$. This is
already position independent.

The leading $1/2$ represents the vacuum energy. To obtain a finite
expression we subtract the expectation value at zero temperature.
\begin{equation}
  \exv{T_{\mu\nu}}_{ren} = \int\limits_p^+ D_{\mu\nu}\K(p)\,\left[
    n(p_0)\rh(p) + \frac12 \delta\rh(p)\right],
\end{equation}
where $\delta\rh(p) = \rh(p)-\rh_0(p)$, and $\rh_0$ is the zero
temperature spectral function.

In a rotationally invariant system only the diagonal elements
survive.  The $00$ component is the energy density
\begin{equation}
  \label{eq:e}
  \ep = \int\limits_p^+ D\K(p)\,\left[ n(p_0)\rh(p) +
    \frac12\delta\rh(p)\right],
\end{equation}
where $D\K(p)\equiv D_{00}\K(p)$. 

We note that $\ep$ is independent on the normalization of the spectral
function (if $\rh\to Z\rh$ then $G_R\to ZG_R$ and $\K\to \K/Z$, so $Z$
drops out). This means that only the energy levels count, not the
normalization (which is the consequence of the definition of the
fields).

Having $\ep(T)$ we can obtain the free energy from the relation $\ep =
\frac{\d \beta f}{\d\beta}$, the pressure from $p=-f$ and the entropy
density from $s=\beta(\ep+p)$. From $\ep$ we can define also a
quantity which can be interpreted as the number of internal degrees of
freedom. At high temperatures $n(p_0)\sim T/p_0$, and the average
energy per mode is $T$. Therefore the number of modes for a given
momentum is
\begin{equation}
  \label{np}
  N_{dof} = \int\limits_0^\infty \frac{dp_0}{2\pi}\, \frac1{p_0}
  D\K(p)\,\rh(p).
\end{equation}
This is similar to the ``equivalent photon number'' introduced by
Weiz\"sacker and Williams.  $N_{dof}$ is dimensionless, temperature
independent quantity which, as we will see below, in case of the
discrete spectrum yields indeed the number of energy levels.

After we have derived the relevant formulae we can apply them for some
particular spectral functions. First of all one easily verifies that
for a relativistic free particle (with spectral function $\rh(p) =
2\pi\sgn p_0 \delta(p^2-m^2)$) one obtains one degree of freedom
($N_{dof}=1$) from \eqref{np}, and the energy density from
\eqref{eq:e} yields the standard formula
\begin{eqnarray}
  \label{eq:ep1}
  \ep_1(m,T)= \frac1{2\pi^2} \int\limits_m^\infty dp_0
  n(p_0)\,p_0^2\sqrt{p_0^2-m^2}.
\end{eqnarray}

Consider now two stable particles with masses $m_{1,2}$ and wave
function renormalizations $Z_{1,2}$. Since the result is independent
on the global normalization of $\rh$, we can choose $Z_1+Z_2=1$ and
consider
\begin{equation}
  \rh(p) = (2\pi)\sgn p_0\left[ Z_1\delta(p^2-m_1^2) + Z_2
    \delta(p^2-m_2^2)\right].
\end{equation}
This yields $G_R(p)= Z_1(p^2-m_1^2)^{-1}+Z_2(p^2-m_2^2)^{-1}$ with
Landau prescription, and $\K(p) = (p^2-m_1^2)(p^2-m_2^2)/(p^2 -\bar
m^2)$ where $\bar m^2 = Z_2m_1^2+Z_1m_2^2$. Then we find
\begin{equation}
  \frac{\d\K}{\d p^2} = \frac{p^4-2p^2\bar m^2+
    Z_2m_1^4+Z_1m_2^4}{(p^2-\bar m^2)^2}.
\end{equation}
If we evaluate the internal degrees of freedom, we obtain $N_{dof}=2$
independently on the details, in particular on the weights of the
Dirac-deltas. If, however, $m_1=m_2$ at the beginning, then (like in
the previous case), we have $N_{dof}=1$. Therefore we indeed count the
number of the particle species, which is discontinuous: this is the
manifestation of the Gibbs paradox in our expression.

Similarly, the energy density reads
\begin{equation}
  \ep_2 = \left\{
    \begin{array}[c]{ll}
      \ep_1(m_1,T)+\ep_1(m_2,T)\quad&\mathrm{if}\; m_1\neq m_2\cr
      \ep_1(m,T),\qquad&\mathrm{if}\; m_1=m_2=m,\cr
    \end{array}\right.
\end{equation}
where $\ep_1$ comes from \eqref{eq:ep1}. This formula is also
non-analytic for $m_2\to m_1$, and independent on $Z_{1,2}$.

To have a deeper insight into the calculations, we remark that in
the energy density \eqref{np} we have the combination $D\K\rh$. Since
$\K=0$ on the mass shells, we have to consider the $\d\K/\d p^2$
term. Near the mass shell it has the following property:
\begin{equation}
  \lim_{p^2\to m_1^2} \left[\lim_{m_2\to m_1} \frac{\d\K}{\d
      p^2}\right] = 1,\quad \lim_{m_2\to m_1} \left[ \lim_{p^2\to
      m_1^2} \frac{\d\K}{\d p^2}\right] = \frac 1{Z_1}.
\end{equation}
It is this non-interchangeability of the limits which lies behind the
Gibbs paradox. A physical interpretation of these limits is that a
finite width gives finite resolution of the spectra; if the peaks are
closer than their widths, than then one cannot resolve them, and we
are in the first limiting case. If they are farther, than we see two
peaks, and the second limiting procedure has to be applied.

The above observation generalizes to any number ($N$) of Dirac-deltas. Let
us take in general
\begin{equation}
  \rh(p_0) = \sum\limits_i Z_i(2\pi) \delta(p_0-\omega_i),\quad G_R
  = \sum\limits_i \frac{Z_i}{p_0-\omega_i+i\epsilon}.
\end{equation}
The kernel is its inverse, the derivative of the kernel reads
\begin{equation}
  p_0\frac{\d\K}{\d p_0} = -\frac{p_0}{G_R^2} \frac{\d G_R}{\d p_0} =
  \frac{p_0}{G_R^2}  \sum\limits_i \frac{Z_i}{(p_0-\omega_i)^2}.
\end{equation}
Since at $p_0\to\omega_i$ for any $i$ the retarded Greens function
diverges, its inverse is zero: ie. $\K(p)=0$ on any mass shell. On the
other hand, if $Z_i\neq0$
\begin{eqnarray}
  \lim_{p_0\to\omega_i} p_0\frac{\d\K}{\d p_0} =&&  \lim_{p_0\to\omega_i} 
  \frac{\omega_i}{Z_i^2/(p_0-\omega_i)^2+\mathrm{finite}}\times\nn&&\times
  \left[\frac{Z_i}{(p_0-\omega_i)^2} + \mathrm{finite} \right] =
  \frac{\omega_i}{Z_i},
\end{eqnarray}
and so the $Z_i$ factors drop out from the result. The number of
degrees of freedom from \eqref{np} is $N_{dof} = N$,
and the energy density is $\ep_N=\sum_n \ep_1(m_n,T)$, both are
independent on the normalization.

After the infinite lifetime examples we consider the case of realistic
gases. First let us consider a single degree of freedom with finite
lifetime, represented by a spectral function of Breit-Wigner type
\begin{equation}
  \rh_1(p,\Gamma) =
  \frac{4p_0\Gamma}{(p_0^2-\omega_p^2-\Gamma^2)^2+4p_0^2\Gamma^2},
\end{equation}
where $\omega_p^2=\p^2+m^2$. The corresponding retarded Greens
function and the kernel are $G_R^{-1}(p) = (p_0+i\Gamma)^2-\omega_p^2$
and $\K(p) = p^2-m^2-\Gamma^2$. We can easily compute the number of
degrees of freedom from \eqref{np} to obtain $N_{dof}=1$! This means
that, independently on the lifetime of the excitations, we always have
a single degree of freedom.

\begin{figure}[htbp]
  \centering
  \hbox{\hspace*{-2em}\includegraphics[height=3.8cm]{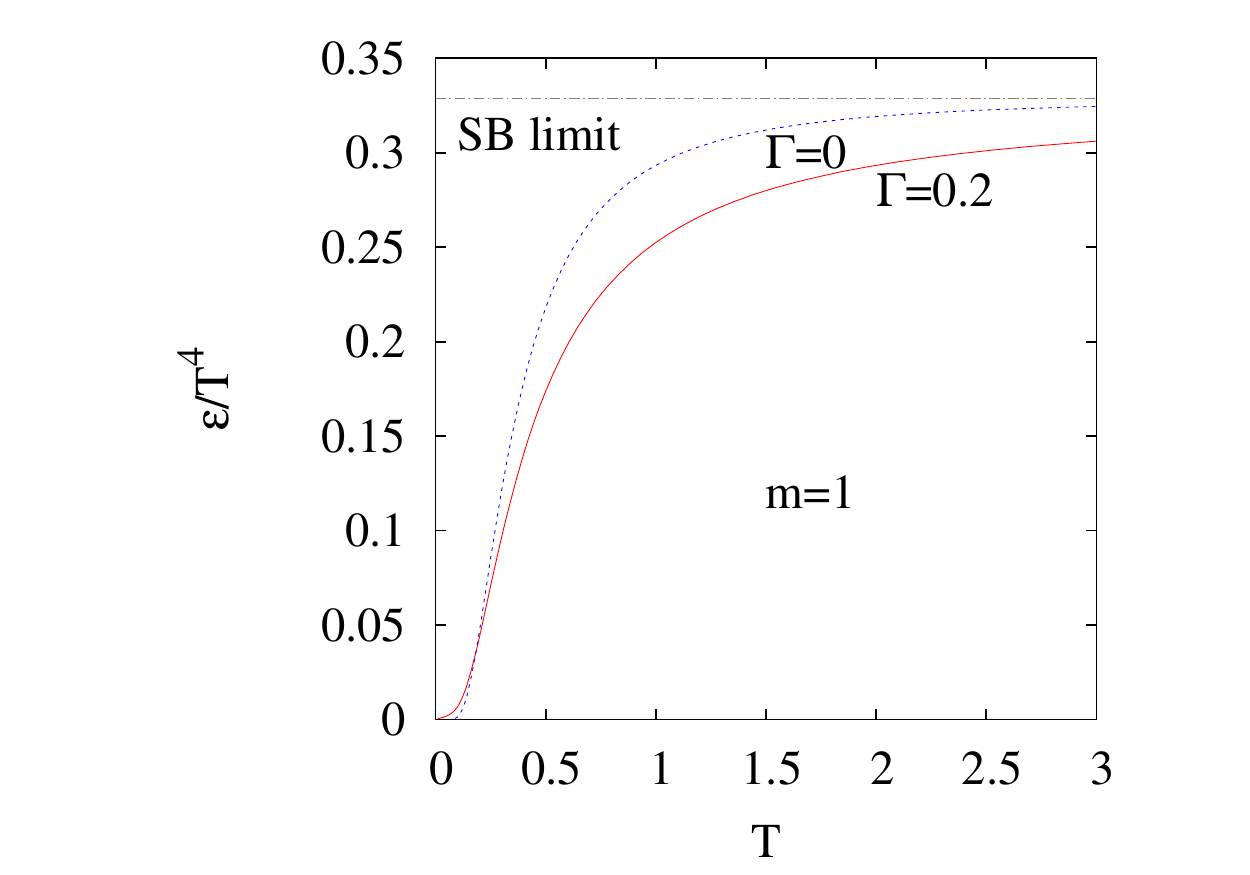}
  \hspace*{-4em}\includegraphics[height=3.8cm]{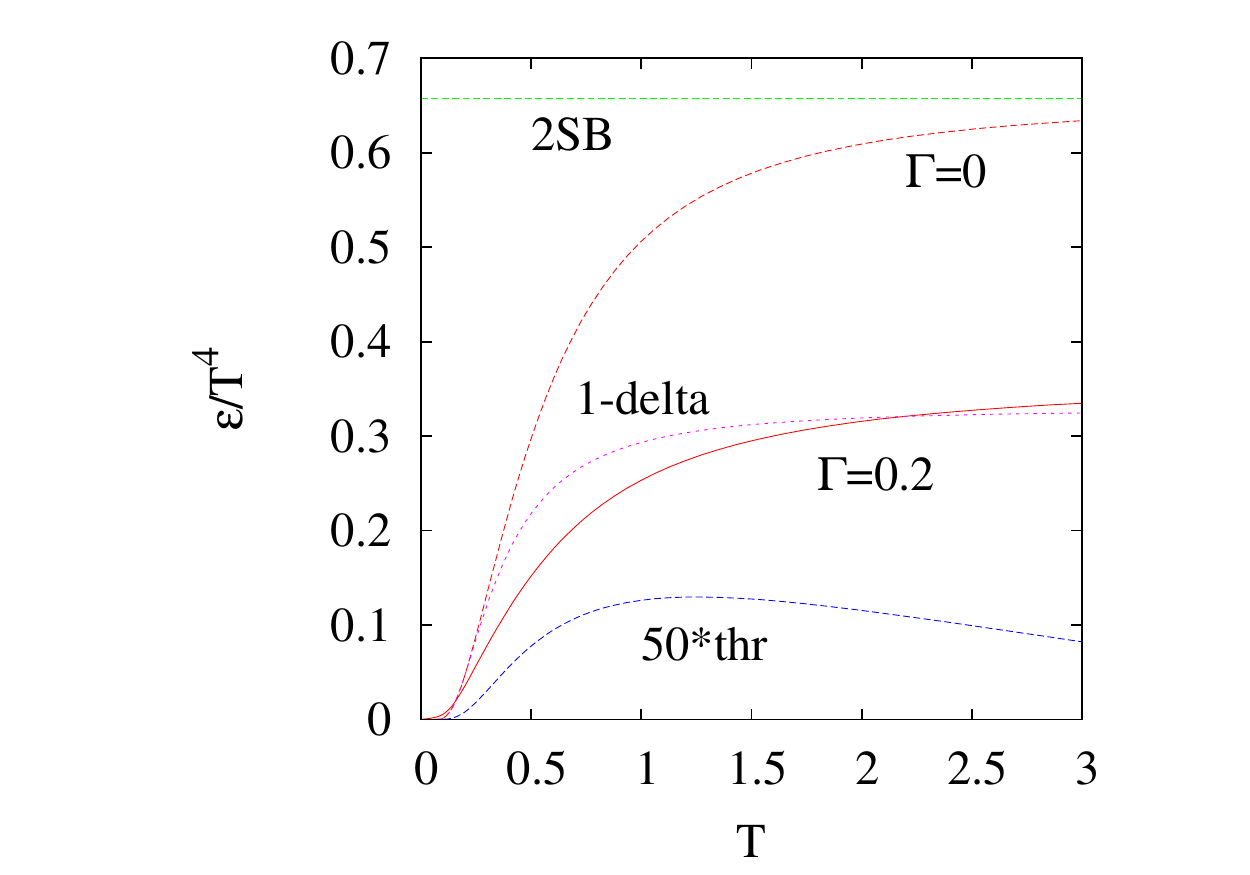}}
  \hbox{\hspace*{7em}a.)\hspace*{12em}b.)}
  \caption{a.) Energy density coming from a single Lorentzian for
    $\Gamma=0$ (ie. infinite lifetime) and $\Gamma/m=0.2$. SB limit
    denotes the Stefan-Boltzmann limit. b.) Energy density for two
    Lorentzians with $m_1=1$ and $m_2=2$ masses and
    $\Gamma_1=\Gamma_2=\Gamma$ width. 2SB means double of the SB
    limit. 1-delta is a \emph{single} Dirac delta case. 50*thr shows
    the 50-times energy density coming from a threshold. }
  \label{fig:Lorentz}
\end{figure}

For the energy density we evaluate \eqref{eq:e}, the result can be
seen on Fig.~\ref{fig:Lorentz}/a.
One can see that for $\Gamma/m$ as large as $0.2$, the energy density
is close to the one for a stable particle. This observation makes
possible to neglect the width for independent real quasiparticles in
thermodynamical calculations.

Although a single Lorentzian yields $N_{dof}=1$, this does not mean --
because of nonlinearity of the expressions \eqref{eq:e} and \eqref{np}
-- that for $N$ Lorentzians we obtain $N$ degrees of freedom. To see
this, let us take the linear combination of two such Lorentzians:
\begin{equation}
  \rh_2(p) =Z_1\rh_1(p,\Gamma_1) + Z_2\rh_1(p,\Gamma_2).
\end{equation}
Then, for $N_{dof}$ we obtain Fig.~\ref{fig:twostates}/b which we also
discussed in the introduction. If the width is small as compared to
the distance of the peaks then we get $N_{dof}=2$: in fact the
$\Gamma\to0$ limit is the case of two Dirac-deltas seen above. If the
width gets larger, the two Lorentzians start to overlap, and we
\emph{continuously} arrive to the case of one broad Lorentzian seen
above with $N_{dof}=1$. The phenomena seen on
Fig.~\ref{fig:twostates}/b. is nothing else than the Gibbs-paradox,
smeared out for real gases.

The energy density can be seen on Fig.~\ref{fig:Lorentz}/b. What we
see is that for $m_1=1$ and $m_2=2$ a particle width $\Gamma=0.2$
is already very close to the 1-particle case. We therefore see the
reduction of degrees of freedom also in the energy density.

Finally we can take a pure threshold with $\rh(p)=
\sqrt{1-m^2/p^2}$. With $m=1$ we see the result for the energy density
on Fig.~\ref{fig:Lorentz}/b. In order to see something at all, we
multiplied the result by 50 -- this suggests that thresholds have very
tiny contribution to the energy density. This result can be
interpreted as the fate of multiple bound states: when they are melted
to a broad continuum, then their contribution vanishes from the EoS.

As a conclusion we stress that one has to be cautious in treating
effective quasiparticle models, even in the weakly interacting
case. Since quasiparticles are composite objects, they do not represent
independent degrees of freedom, only in the limit when their mass
difference is much larger than their width. As a consequence their
contribution to different physical quantities like energy density do
not simply add up: we observe nonlinear behavior, when two finite
width quasiparticles can melt into a single one, or multiparticle
states can give vanishing contribution
(cf. Fig.~\ref{fig:Lorentz}). We expect that this nonlinear behavior
may play important role also in other physical effects, like charm
suppression or transport phenomena in fluids. Our method can also be
used to identify other currents and compute transport coefficients
from the effective theory. This is the subject of ongoing projects.

\begin{acknowledgments}
  The author thanks useful discussions with T.S. B\'{\i}r\'o,
  G. Gy\"orgyi, I. Nagy, A. Patk\'os, Z. R\'acz and Zs. Sz\'ep. This
  work is supported by the Hungarian Research Fund (OTKA) under
  contract No. K68108.
\end{acknowledgments}

\end{document}